\documentclass[rnote]{aa} % for the research notes
\usepackage{graphicx}
\usepackage{txfonts}
\usepackage{natbib}            % pour utiliser bibtex ...
\bibpunct{(}{)}{;}{a}{}{,} % to follow the A&A style
\usepackage{epsfig}
\usepackage{lscape}
\begin{document}
   \title{Search for cold gas along radio lobes in the cooling core galaxies MS0735.6+7421 and M87}

   \author{P. Salom\'e
          \inst{1}
          \and
          F. Combes\inst{2}}

   \offprints{P. Salom\'e}

   \institute{Institut de Radio Astronomie Millim\'etrique  (IRAM), Domaine
     Universitaire, 300 rue de la piscine. F-38400 St Martin d'H\`eres, France\\
              \email{salome@iram.fr}
         \and
             LERMA, Observatoire de Paris, 61 av. de l'Observatoire, 
	     F-75014 Paris, France 
	   }

   \date{Received May xx, 2008; accepted May xx, 2008}

  \abstract{We report CO observations towards MS0735.6+7421 a distant
  cooling core galaxy, and towards M87, the nearest cooling core in
  the center of the Virgo cluster.  Both galaxies contain radio
  cavities that are thought to be responsible for the heating that can
  regulate or stop the cooling of the surrounding gas. In this
  feedback process, there could still be some gas cooling along
  filaments, along the borders of the radio cavities.  Molecular gas
  is known to exist in clusters with cooling cores, in long and thin
  filaments that can be formed behind the rising bubbles inflated by
  the central AGN.  CO emission was searched for at several locations
  along the radio lobes of those two galaxies, but only upper limits
  were found. These correspond to cold gas mass limits of a few
  10$^9$M$_\odot$ for each pointing in MS0735.6+7421, and a few
  10$^6$M$_\odot$ in M87.  This non detection means that either the
  cooling is strongly reduced by the AGN feedback or that the gas is
  cooling in very localized places like thin filaments, possibly
  diluted in the large beam for MS0735.6+7421. For M87, the AGN
  heating appears to have stopped the cooling completely.

   }

   \maketitle

%---------------------------------------------------------------------------
\section{Introduction}
%---------------------------------------------------------------------------

Giant ellipticals in galaxy cluster centers live in a complex
environment. In some cluster cores the intra-cluster medium is known
to cool down, but how much of this gas is able to fuel the central
galaxy is still not clear.  The hot gas is no longer thought to be
flowing directly towards the central galaxy.  The regulation of the
intra-cluster medium (ICM) cooling involves some re-heating mechanisms
like jet feedback from a central AGN and it becomes observable by the
formation of X-ray emitting gas cavities. This feedback mechanism
produces mass outflows, weak shocks and heating \citep{McNa2007}, but
also cooling, in long and thin filaments.  Indeed, some very cold gas
(10-100 K) has been observed in CO emission in many cooling core
systems \citep{Edge01, Salo03}.  High resolution imaging of CO showed
the existence of molecular filaments in Abell 1795 \citep{Salo04} that
are confined to the borders of the radio cavities. These filaments
extend even farther outward in NGC~1275 \citep{Salo08}. The cold gas
formation may result from the interaction of the AGN with the
surrounding hot gas. The filaments are thought to be cooling filaments
that can fuel the giant elliptical with cooled gas.  More examples of
this phenomenon need to be observed to better understand the physics
of the formation of the cold gas. In this work, we report a search
towards two cooling core systems: MS0735.6+7421 and M87 in the Virgo
cluster. These two objects are known to be the sites of very strong
feedback. The cooling in M0735.6+7421 very likely occurs only in the
very central region and M87 is a poor cooling core with almost no
cooling, so we do not expect to find much cooled gas.

%%%%%   M0735 intro

\begin{figure}[htbp]
\centering
\epsfig{file=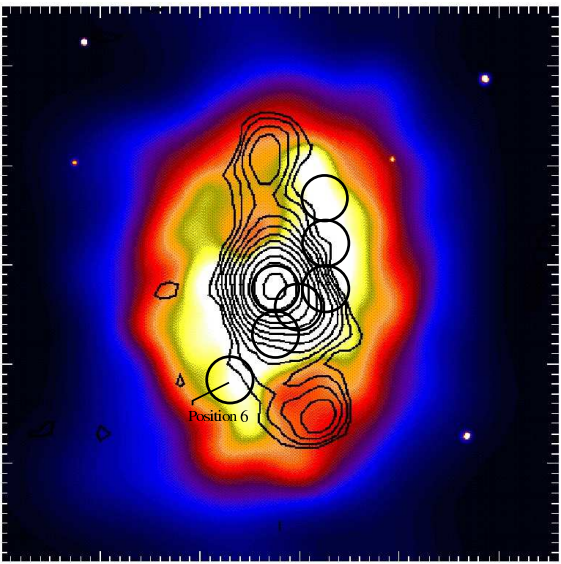, width=7.5cm}% &
\caption{{A 250 $\times$ 250$''$ image of the
MS0735.6+7421 cluster center \citep{McNa05}, one arcsec is 3.5
kpc. The 1.4 GHz radio contours of the jets emitted by 4C+74.13 are
superposed. The cavities in the X-ray emission coincide with the radio
lobe positions (200 kpc in size at the cluster redshift). The circles
represent the 30m telescope beam at 3mm, showing the 7 regions we
observed in CO(1--0).  }}
\label{over}
\end{figure}

The giant galaxy MS0735.6+7421 is at a distance of 716 Mpc
(1$''$ correponds to 3.47 kpc). In spite of its large distance, its
giant X-ray cavities are so large, inflated by radio lobes of size 550
kpc, that they are resolved by the beam of the IRAM-30m telescope.  It
is then possible to probe the cold gas distribution and compare it
with the X-ray, H$\alpha$ and radio maps.  The galaxy MS0735.6+7421
hosts a radio source 4C+74.13.
According to \citep{McNa05} the AGN activity could have re-heated
sufficiently the ICM to stop the cooling around the central
galaxy. Figure \ref{over} shows that the radio lobe emission of
4C+74.13 fills the X-ray cavities of the intra-cluster medium. These
cavities may have displaced and compressed the hot gas surrounding the
central cluster galaxy, as predicted by feedback models of cooling
flows. Negative feedback models as described by \citet{Revaz08}
predict that the hot X-ray gas when compressed may eventually cool
very efficiently down to very low temperatures, forming cold gas
filaments visible in CO(1--0) emission. So in this galaxy, we searched
for molecular gas emission not only in the central region but also far
away from the X-ray cooling core region, along the AGN radio lobes.

%%%%%   M87 intro

M87 is the most nearby X-ray bright cooling core. It lies at a
distance of 18 Mpc.  X-ray mapping around M87 reveals a complex
structure for the cooling gas, perturbed by several bubbles (Young et
al 2002).  The X-ray-cool component detected in the radio lobes is in
blobs that are connected to the radio buoyant bubbles
\citep{Ghizzardi2004}.  In a deeper XMM-Newton observation of M87,
\citet{Simio2007} find evidence that cold, metal-rich gas is being
transported outwards also outside the X-ray arms, possibly through
bubble-induced mixing.  Many X-ray features are better be understood
if there are repetitive AGN outbursts.  Depressions appear as the
remnants of earlier outbursts, while shocks are seen as brightenings
within the prominent X-ray arms.  As in Perseus, shocks may be the
most significant channel for AGN energy input into the cooling-flow
atmospheres, and outbursts every 3$\times$10$^7$ yr are sufficient to
quench the cooling \citep{Forman2005}.

%%%%%%%%%%%%%%%%%%%%%%%%%%%%%%%%%%%%
%  OFFSET POSITIONS OVER X_RAY MAP
%%%%%%%%%%%%%%%%%%%%%%%%%%%%%%%%%%%%

%----------------------------------------------------------------------- 
\section{Observations}
%----------------------------------------------------------------------- 
%
The observations were made with the IRAM 30m telescope on Pico Veleta,
Spain during two different runs in summer 2005 and summer
2006. Weather conditions were not optimal. The first run (august 2005)
suffered from wind and significant pointing corrections. The second
run (may 2006) was done with clouds and 10-15\,mm \, of precipitable water
vapor, that means higher system temperatures.  We used the wobbler
switching mode with two 3mm and two 1.3mm receivers operating
simultaneously.  The beam throw was 120$''$. For MS0735.6+7421,
receivers were tuned to the CO(1--0) and HCN(2--1) emission lines,
redshifted to the velocity of the source. The CO(2--1) emission line
is not visible from the 30m telescope at this redshift ($z=0.216$).
For M87, we observed simultaneously the CO(1--0) and CO(2--1) ines, at
3mm and 1.3mm.  Frequent pointings and focus checks were carried out
using bright quasars.  At 3mm, we used two 512\,$\times\,$1\,MHz
filterbanks. This gives a total band of $\sim$ 1300\,km/s for the
CO(1--0) line setup.  In addition, we used the two 250\,$\times\,$\,4
MHz resolution filterbanks for the 1.3mm receivers. The beams of the
30m telescope, at 3mm and 1.3mm, are 28$^{\prime\prime}$ (94.795 GHz)
and 13$^{\prime\prime}$ (218.656 GHz) respectively for MS0735.6+7421
and 23$^{\prime\prime}$ (114.771 GHz) and 12$^{\prime\prime}$ (229.537
GHz) for M87.

The signals are expressed here in main beam brightness temperatures.
The ratio of the antenna temperature to the main beam temperature is 
${\rm T_{\rm A}^*/T_{\rm mb} = B_{eff}/F_{eff}}$ with the ratio of the
main-beam efficiency to the forward efficiency being ${\rm
B_{eff}/F_{eff}}$= 0.75/0.95 at 3mm and 0.52/0.91 at 1.3mm (cf
IRAM-30m site http://www.iram.es/).

The data were calibrated with the MIRA software and reduced with the
CLASS90 package. Spiky channels and bad scans were dropped and linear
baselines were subtracted for each spectrum.  After averaging all the
spectra for each line at each position, the data were Hanning smoothed
The results are summarized in Tables \ref{table-upper} and \ref{table-upper-m87}.
%

%---------------------------------------------------------------------------
\section{Results} 
%---------------------------------------------------------------------------

We expected the regions where the gas can cool efficiently to be along
the radio lobes and X-ray cavities, as in Abell 1795
\citep{Salo04}. This is where we looked for a past or still present
cooling activity with CO line observations. In MS0735.6+7421, we
observed 7 fields to search for CO(1--0) emission in the brightest
X-ray regions in the radio lobes. In M87, we observed also towards 4
offsets on the borders of the radio lobes. Only upper limits have been
obtained in both galaxies.

For MS0735.6+7421 at a redshift of $z=0.216$, assuming a Hubble
constant of 71 km/s/Mpc, and a flat universe with cosmological
parameters $\Omega_m= 0.27$ and $\Omega_\Lambda= 0.73$, the luminosity
distance of MS0735.6+7421 is D$_L$ = 1060 Mpc.  The CO luminosity L$^\prime_{\rm CO}$ 
[$\rm K\,km\,s^{-1}\,pc^{-2}$] is given by \cite{Solo97}:
$$ L^\prime_{\rm CO} = 3.25\,10^7 \left[\frac{S_{\rm CO} \Delta\,V}{\rm
Jy\,km.s}\right] \left[{{\nu_{\rm obs}}\over {\rm GHz}}\right]^{-2}
\left[{{D_L}\over {\rm Mpc}}\right]^2 (1+z)^{-3}$$ We computed H$_2$
masses by using the standard conversion factor $M_{\rm H_2} = 4.6\,
L^\prime_{\rm CO}$ M$_\odot$ that assumes self-gravitating clouds. Note that
this mass could be overestimated by a factor 5 if the gas is optically
thin and subthermally excited \citep{Downes98}. We applied the same formula to determine
the CO luminosity of M87 and assumed a distance of 18 Mpc (z=0.00436) to this
source.
The integrated CO intensity   S$_{\rm CO}$ [Jy\,km/s] is:
\begin{equation}
{\rm S}_{\rm CO}= 4.95 \int T_{\rm mb}{\rm (CO)}\, dV \,,
\end{equation}  
where $T_{\rm mb}{\rm (CO)}$ is the main beam brightness
temperature, obtained with the CO emission line, $\rm dV$ is the line
width and 4.95 the telescope S/T$_{\rm mb}$ factor in Jy per K. 
To express upper limits, we smoothed the spectra to 50 km/s resolution,
and assumed a standard linewidth of $\rm dV$ = 300\,km/s.

%---------------------------------------------------------------------------
\subsection{MS0735.6+7421}
%---------------------------------------------------------------------------

The analysis of the data does not show any evidence of CO(1--0) in any
of the 7 positions investigated.

%%%%%%%%%%%%%%%%%%%%%%%%%%%%%%%%%%%%%%%%%%%%%
% RESULTS - UPPER LIMITS
%%%%%%%%%%%%%%%%%%%%%%%%%%%%%%%%%%%%%%%%%%%%%
\begin{table}[h]
\centering
\begin{tabular}{cccccc}
\hline
\hline
Offset  ($\alpha$,$\delta$)   &   Sigma & I$_{\rm CO}$ & Time  &  T$_{\rm sys}$  & M$_{\rm gas}$\\
 $[\rm{arcsec}]$&    [mK] & [Kkm/s] &  [min] &   [K] &  [10$^9$ M$_\odot$]\\
\hline
(0 $\times$ 0)  & 0.8 & 0.24 & 90 & 137  & $\le$  12.4 \\
(-22 $\times$ 0)  & 0.9 & 0.27 & 90 & 135  & $\le$  13.9 \\
(-22 $\times$ 22)  & 0.4 &0.12 &  210 & 146  & $\le$  6.2 \\
(-11 $\times$ -11)  & 0.8 & 0.24 & 66 & 131  & $\le$  12.4 \\
(0 $\times$ -22)  & 0.4 & 0.12 & 174 & 135  & $\le$  6.2 \\
(22 $\times$ -44)  & 0.3 &0.09 &  573 & 189  & $\le$  4.6 \\
(-22 $\times$ 44)  & 1.9 & 0.57 & 31 & 209  & $\le$  29.3$^1$ \\
\hline
\end{tabular}
\caption{Summary of CO(1--0) observations, sensitivity and emission
upper limits for MS0735.6+7421. All linewidths have been assumed to be
300\,km/s.  The mass upper limits are 1$\sigma$ limits.  Offsets in
arcsec are from the central position J2000 RA 07:41:44.5, DEC
+74:14:40. $^1$Note that the last offset was observed during only 31
min. This explains the larger upper limit we obtain for that region.}
\label{table-upper}
\end{table}
%%%%%%%%%%%%%%%%%%%%%%%%%%%%%%%%%%%%%%%%

At the distance of MS0735.6+7421, the CO(1--0) spectra could be very
broad (due to the mixing of many molecular clouds at different
velocities) and even broader in the very central region \citep{Salo08} 
making it very difficult to detect. Stacking the emission of
all the offsets together did not improve much the signal to noise. This
means that either the amount of cold gas is small, or a large velocity
dispersion exists.

For MS0735.6+7421, the 28$''$ field of view of the 30m telescope
corresponds to a diameter of 97 kpc. Recent observations of NGC~1275
\citep{Salo08} have shown that the molecular gas, in cooling core
galaxies, may lie in very thin and elongated filaments of at most
$\sim$(2 kpc $\times$ 10 kpc).  If this is the case for MS0735.6+7421,
it means a beam filling factor of 0.4$\%$ for such a filament. So even
with a large collection of filaments, we expect a significant dilution
effect in the antenna beam.

The upper limits on the amount of cold molecular gas are summarized in
Table \ref{table-upper}. We find on average M$_{\rm gas} \le$ a few
10$^9$M$_{\odot}$ inside a beam of diameter 97 kpc.  In a multiphase
cooling flow scenario, the mass deposition rate evaluated from X-rays
is 260 M$_\odot/yr$ inside 30$''$ \citep{Gitti2007}. For a typical
cooling time of 10$^8$yr, this means that a steady reservoir of at
least 10$^{10}$M$_\odot$ would exist (and even more for a star
formation rate lower than the mass deposition rate). Our total upper
limit, obtained by summing over all the central regions observed, is
close to a few 10$^{10}$M$\odot$. Considering the uncertainties on the
mass evaluation, this means that we cannot rule out that some gas is
indeed cooling but not yet visible in CO.

\citet{Edge01} and \citet{Salo03} have shown the existence of a
correlation between CO emission and H$\alpha$ luminosity in cooling
core clusters. If we refer to this relation, we can evaluate the
expected molecular gas mass from the known H$_\alpha$ luminosity
coming from the central 25 kpc \citep{Donahue1992}. This gives$\sim$
6$\times$10$^{10}$ M$_\odot$ which is slightly larger than, but still
consistent with, the 1$\sigma$ upper limits we found here (offset
[0,0]). We do not find CO line emission either far away from the
cooling cores (other offsets), where there is no optical emission
line.

%---------------------------------------------------------------------------
\subsection{M87}
%---------------------------------------------------------------------------

%%%%%%%%%%%%%%%%%%%%%%%%%%%%%%%%%%
% M87 X-ray and radio Image
%%%%%%%%%%%%%%%%%%%%%%%%%%%%%%%%%%
\begin{figure}[htbp]
\centering
\begin{tabular}{lll}
\includegraphics[width=8cm]{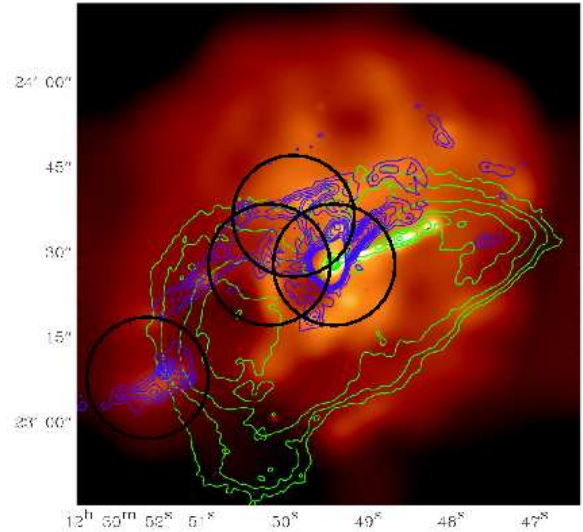} &  
\end{tabular}
\caption{M87 - X-ray image of M87 by \cite{Young02}. The white
contours are the 6cm radio emission and the dark contours show the
H$_\alpha$ nebulosities along the edges of the radio lobes.  The circles
represent the 30m telescope beam size at 3mm.}
\label{over-m87}
\end{figure}
%%%%%%%%%%%%%%%%%%%%%%%%%%%%%%%%%%%

%%%%%%%%%%%%%%%%%%%%%%%%%%%%%%%%%%%%%
% RESULTS - UPPER LIMITS M87
%%%%%%%%%%%%%%%%%%%%%%%%%%%%%%%%%%%%%
\begin{table}[htbp]
\centering
\begin{tabular}{lccccc}
\hline
\hline
Line & Offset  ($\alpha$,$\delta$)  &   Sigma & Time  &  T$_{\rm sys}$  & M$_{\rm gas}$\\
 & [arcsec]  & [mK] &  [min] &   [K] &  [10$^6$ M$_\odot$]\\
\hline
CO(1--0) &(0 $\times$ 0)  & 1.8 &  318 & 281  &  $\le$  9.7 \\
CO(2--1) &(0 $\times$ 0)  & 1.8 &  318 & 680  &   \\
CO(1--0) &(5.5 $\times$ 11) & 1.6 &  258 & 371  &  $\le$  8.6 \\
CO(2--1) &(5.5 $\times$ 11) & 2.2 &  258 & 1020  &   \\
CO(1--0) &(11 $\times$ 0)  & 1.7 &  361.5 & 264  & $\le$  9.2 \\
CO(2--1) &(11 $\times$ 0)  & 2.5 &  349.5 & 465  &     \\
CO(1--0) &(33 $\times$ -22)  & 1.1 &  260 & 330  & $\le$  5.9 \\
CO(2--1) &(33 $\times$ -22)  & 3.2 &  260 & 821  &   \\
\hline
\end{tabular}
\caption{M87 - Summary of CO(1--0) and CO(2-1) observations : sensitivity and emission
upper limits. All linewidths have been assumed at 300km/s.
The mass upper limits are given at 1$\sigma$.}
\label{table-upper-m87}
\end{table}
%%%%%%%%%%%%%%%%%%%%%%%%%%%%%%%%%%%%%%%

M87 has been identified as a very moderate cooling core \citep{Forman2007}. 
It is possible that the present AGN heating has completely quenched
the cooling flow.  \cite{Young02} compared the optical emission with
Chandra X-ray images (see Fig.\ref{over-m87}).  Filaments are observed
along the edges of the radio lobes, which coincide with the sites where the X-ray
gas is cooler, as in usual cooling core clusters. The filaments
have a total H$\alpha$ luminosity of 3$\times$10$^{40}$erg.s$^{-1}$ in 1250
arcsec$^2$, \citep{Sparks93}.

Dynamical models of hot bubbles of gas injected by the active galactic
 nuclei (AGN), rising buoyantly into the intra-cluster volume, indicate
 that the activity cycle of the AGN in M87 is roughly 10$^8$ yr
 \citep{Kaiser2003}.  The largest radio structure is likely to be the
 remnant of an old bubble \citep{Mathews2008}. X-ray observations show
 a 30 kpc-long nearly radial filament of relatively cooler gas,
 aligned in projection with the radio lobe.

We have observed 4 points aligned along the radio lobes and H$\alpha$
nebulosities (see Figure \ref{over-m87}). Only upper limits are found
(see Table \ref{table-upper-m87}).  The center of the galaxy is also
not detected, see also NGC 4486 in \cite{Combes2007}.  This suggests
that the feedback heating is very likely to be efficient enough to suppress
the cooling, otherwise we should be able to observe cold gas falling
back towards the center.

%---------------------------------------------------------------------------
\section{Discussion and conclusions}
%---------------------------------------------------------------------------

We observed 7 different positions along the radio lobes of
MS0735.6+7421 and 4 positions for M87, in CO(1--0) with the 30m telescope. 
We do not detect cold molecular gas in any of the observed
regions.  Our upper limits put significant constraints on the
cold gas masses. We find M$_{\rm gas}\,\le$ a few
10$^9$M$_\odot$ in MS0735.6+7421 and M$_{\rm gas}\,\le$ a few
10$^6$M$_\odot$ in M87. The absence of cold gas could be explained if strong
feedback heats sufficiently the intra-cluster medium and prevents any
cooling. Nevertheless, as suggested by the recent results of high
resolution CO mapping of NGC~1275 \citep{Salo08}, if the cooling
happens only in very thin and elongated filaments, we may have missed
such emission in MS0735.6+7421. The spatial and spectral dilution of a collection of CO
filaments would make the detection difficult in a
single beam of 28$''$ (97 kpc) at this distance. 
 More sensitive observations, with higher spatial resolution
are needed to determine whether this is the case.

%%%%%%%%%%%%%%%%%%%%%%%%%%%%%%%%%%%%%%%%%%%%%%%%%%%%%%%%%%%%%%%%%%%%%%%%%%%%%%%%%%%%%
% CO(1--0) SPECTRA 
%%%%%%%%%%%%%%%%%%%%%%%%%%%%%%%%%%%%%%%%%%%%%%%%%%%%%%%%%%%%%%%%%%%%%%%%%%%%%%%%%%%%%
\begin{figure}[htbp]
\centering
\includegraphics[width=8.1cm, angle=0]{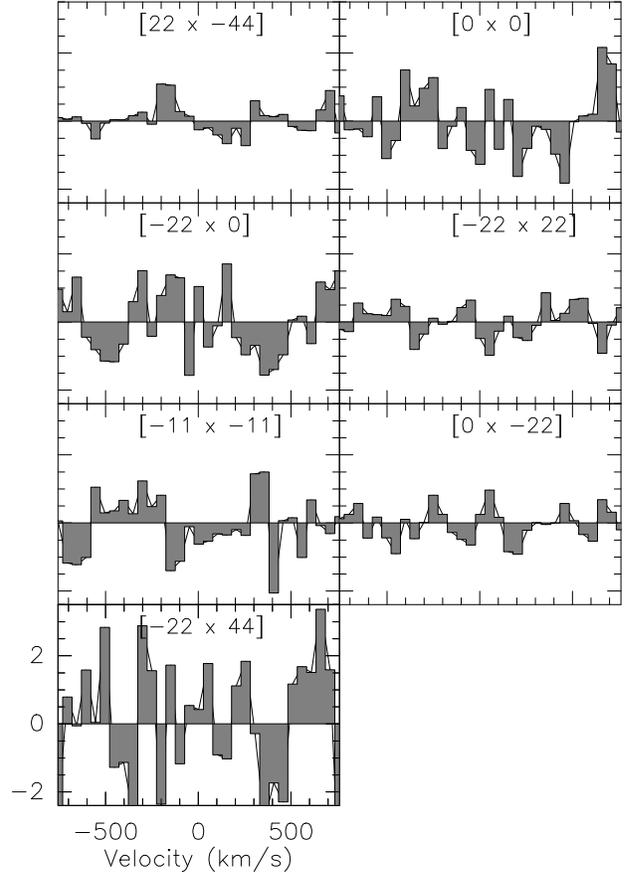}
\caption{MS0735.6+7421 - CO(1--0) spectra obtained at all the positions observed as
indicated at upper right in each diagram and as overlaid on
Fig. \ref{over}. The channel width is 50.6 km\,s$^{-1}$. See Table
\ref{table-upper} for a summary. The Y-axis is in T$_{\rm mb}$ (mK).}
\label{detection}
\end{figure}

%%%%%%%%%%%%%%%%%%%%%%%%%%%%%%%%%%%%%%%%%%%%%%%%%%%%%%%%%%%%%%%%%%%%%%%%%%%%%%%%%%%%%
% CO(1--0) and CO(2-1) SPECTRA UPPER
%%%%%%%%%%%%%%%%%%%%%%%%%%%%%%%%%%%%%%%%%%%%%%%%%%%%%%%%%%%%%%%%%%%%%%%%%%%%%%%%%%%%%
\begin{figure}[htbp]
\centering
\includegraphics[width=8.1cm, angle=0]{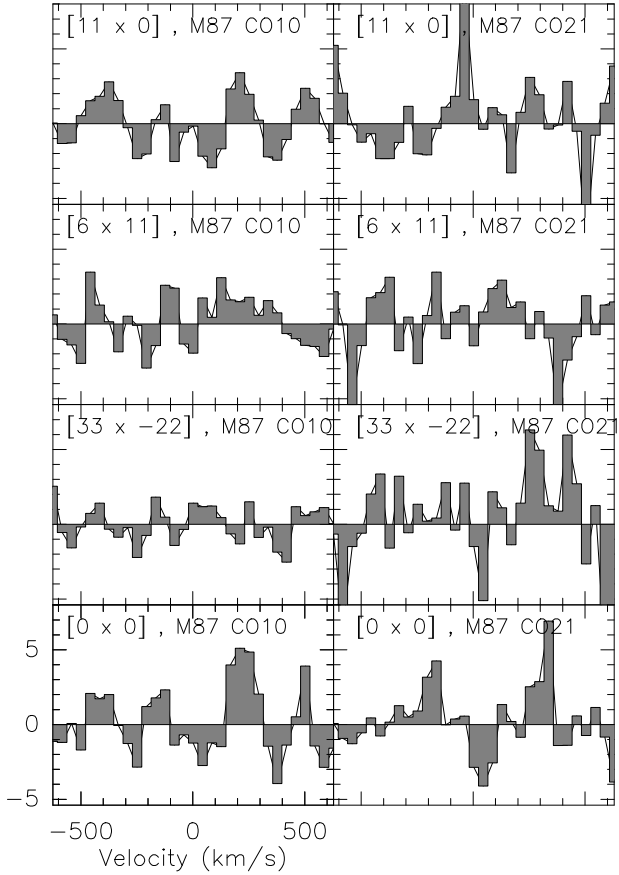}
\caption{M87 - CO(1--0) and CO(2-1) spectra obtained at all the positions observed as
indicated at upper right in each diagram and as overlaid on
Fig. \ref{over-m87}. The channel width is 41.8 km\,s$^{-1}$. See Table
\ref{table-upper-m87} for a summary. The Y-axis is in T$_{\rm mb}$ (mK).}
\label{detection}
\end{figure}
%%%%%%%%%%%%%%%%%%%%%%%%%%%%%%%%%%%%%%%%%%%%%%%%%%%%%%%%%%%%%%%%%%%%%%%%%%%%%%%%%%%%%

%%%%%%%%%%%%%%%%%%%%%%%%%%%%%%%%%%%%%
\begin{acknowledgements}
This work is based on observations done with the IRAM-30m
telescope. IRAM is supported by INSU/CNRS (France), MPG (Germany) and
IGN (Spain). The authors thank the IRAM interferometer staff for help
during the observations.
\end{acknowledgements}
%%%%%%%%%%%%%%%%%%%%%%%%%%%%%%%%%%%%

\bibliographystyle{aa} 
\bibliography{./ref}

\end{document}